\documentclass[showpacs,preprintnumbers,amsmath,amssymb]{revtex4}

\usepackage{graphicx}
\usepackage{dcolumn}
\usepackage{bm}
\usepackage{epsf}
\usepackage[usenames]{color}

\begin{document}

\preprint{APS/123-QED}

\title{Description of $^4$He tetramer bound and scattering states}

\author{Rimantas Lazauskas}
\email{lazauskas@lpsc.in2p3.fr}
\homepage{http://lpscwww.in2p3.fr/theo/Lazauskas/Eng/home.htm}
\affiliation{Groupe de Physique Th\'eorique, Institut de Physique Nucl\'eaire, F-91406 Orsay Cedex,
France.}%

\author{Jaume Carbonell}
\email{carbonell@lpsc.in2p3.fr}
\homepage{http://lpscwww.in2p3.fr/theo/Carbonell/Jaume.html}
\affiliation{Laboratoire de Physique Subatomique et de Cosmologie, 53, avenue des
Martyrs, 38026 Grenoble Cedex, France.}%

\date{\today}

\begin{abstract}
Faddeev-Yakubovski equations are solved numerically for $^4$He
tetramer and trimer states using realistic helium-helium
interaction models. We describe the properties of ground and
excited states, and we discuss with a special emphasis the
$^4$He-$^4$He$_3$ low energy scattering.
\end{abstract}

\maketitle

\section{\label{sec:Intro}Introduction}

Helium atom is one of the simplest few-body systems. Nowadays, its
structure can be described  theoretically with an accuracy
 better than the spectroscopy one~\cite{Atoms}.
In spite of  this simplicity, the compounds of Helium
atoms display a series of unique microscopic and macroscopic
physical phenomena.

The closed shell electronic structure, as well as the compactness
of the Helium atom, makes it the most chemically inert noble gas.
Nevertheless, the very weak van der Waals attraction between two
distant He atoms is responsible for the fact that at very low
temperatures,  both bosonic $^{4}$He and fermionic $^{3}$He
liquify. In addition, the extreme weakness of the He-He
interaction is decisive in explaining why liquid Helium is the
only known superfluid.

As much exciting is the physics of the atomic Helium nanoscopic
structures (He multimers). Recent inspiring
experiments~\cite{Spuch,Toennies} have demonstrated the existence
of bound diatomic $^{4}$He systems (dimers) -  the ground state
with the weakest binding energy of all naturally existing diatomic
molecules. Moreover, diffraction experiment projecting a molecular
beam of small He clusters on nanostructured transmission gratings,
allowed a direct measurement of the dimer average bond length
value $<R>=52\pm 4$ $\AA$. Such a large bond, compared to the
effective range of the He-He potential, enables one to estimate
its binding energy
$E\approx \frac{\hbar ^{2}<R>^{2}}{4m}=1.1+0.3/-0.2$ mK, as well as the $^{4}$He-$%
^{4}$He scattering length  $a_{0}=104+8/-18$ $\AA$.

The  extreme weakness  of the $^{4}$He dimer binding energy
requires a precise theoretical description of the {\it ab initio}
He-He potential,  which results from subtracting the huge energies
of separated atoms. Nevertheless, several very accurate
theoretical
models~\cite{Aziz,Aziz1,Aziz2,Aziz3,Tang_toennies,Aziz4} were
recently constructed, predicting Helium properties in full
agreement with experiments. These effective He-He potentials are
dominated by the strong repulsion (hard-core) at distances
$R_{He-He}\lesssim2$ $\AA$, where the two He atoms are
"overlapping". At larger He-He distances, the weak van-der-Waals
attraction takes over, creating a shallow attractive pocket with a
maximal depth of $V_0\approx 10.9$ K,  centered at
$R_{He-He}\approx3 \AA$. The strong repulsion  of the He-He
potential at short distance,  allows to treat accurately atomic He
systems without having to take into account the internal structure
of single He atoms.

The physics of small He clusters is an outstanding laboratory for testing different
quantum mechanical phenomena. The bound state of $^{4}$He dimer,
at practically zero energy, suggests the possibility of observing
an Efimov-like state in the triatomic compound~\cite{Efimov}.
The existence of this $^{4}$He dimer bound state very
close to threshold, is also responsible for a resonant $^{4}$He-$^{4}$He S-wave
scattering length $a_0 \approx$104 $\AA$, which is more than 30 times larger
than the typical length scale  given by
the Van der Waals interaction $l \sim 3\AA$. The small He clusters low energy scattering
observables should therefore be little sensitive to the details of
He-He interaction and  should thus exhibit some universal
behavior. This sharp separation of different scales can be exploited, making He
multimers a perfect testground for Effective Field Theory (EFT) approaches
~\cite{EFT_intro1,EFT_intro2,Hammer,Platter}.

We present in this work  a rigorous theoretical study of the smallest $^{4}$%
He clusters. There exist in the literature a large number of
theoretical calculations on triatomic $^{4}$He
(trimer)~\cite{Gloeckle,Sandhas,Barletta}. However the four-atomic
$^{4}$He system (tetramer), being by an order of magnitude more
complex in its numerical treatment, remains practically
unexplored. Our study tries to fill up this gap, by providing
original calculations for tetramer bound and scattering states. In
addition, some existing ambiguities in triatomic $^{4}$He
calculations~\cite{Roudnev_ss} are discussed.

To describe $^{4}$He tetramer, Faddeev-Yakubovski (FY) equations
in configuration space are solved. In some cases, this method may
be cumbersome and numerically expensive.  It constitutes
nevertheless a very general and mathematically rigorous tool, with
the big advantage over many other techniques, that it enables a
systematic treatment of  bound and continuum states.

\bigskip

The paper is structured as follows: in
section~\ref{Sec:Theory} we present the theoretical framework used;
in section~\ref{Sec:Trimer} we highlight our $^{4}$He trimer
results. Section~\ref{Sec:tetramer} deals with four-atomic helium
systems and in section~\ref{Sec:rotational}, we discuss
the possible existence of the rotational states in three-atomic and
four-atomic$^{4}$He systems. Section~\ref{Sec:Conclusions} concludes
this work with the final remarks.\bigskip

To describe the interaction between the helium atoms, we have used
the potential developed by Aziz and Slaman~\cite{Aziz}, popularly
referred to as LM2M2 potential. There exists several equivalent
interaction models. However, they all have similar structure and
quantitatively provide very close results~\cite{Barletta,Sandhas_last}.

All results presented in this paper are restricted to the bosonic $^{4}$He
isotope; therefore in the following, we omit the mass number 4 and refer to $%
^{4}$He as He. All calculations use $\frac{\hbar ^{2}}{m}=12.12$ K$\cdot $%
\AA $^{2}$ as the input mass of He atoms.

\section{Theoretical framework\label{Sec:Theory}}

\subsection{Faddeev-Yakubovski equations}

The Schr\"{o}dinger equation is the paradigm of non-relativistic
quantum mechanics. However this equation is not able to separate
different rearrangement channels in the asymptote of multiparticle
wave function (w.f.). Thus it does not provide explicitly a way to
implement the physical boundary conditions for scattering w.f.,
which are necessary to obtain a unique solution.
 Faddeev~\cite{Faddeev} have succeeded to show
that these equations can be reformulated by introducing some
additional physical constraints, what leads to mathematically
rigorous and unique solution of the three-body scattering problem.
Faddeev's pioneering work was followed by Yakubovsky.
In~\cite{Yakubovski}, the systematic generalization of Faddeev
equations for any number of particles was presented.

One should mention that Schr\"{o}dinger equation may still be
applied when solving few-particle bound state problem. However
then one must deal directly with total systems w.f., which is
fully (anti)symmetric and has quite complicated structure.
Exploiting the knowledge of systems symmetry one often tries to
decompose this w.f. into only partially (anti)symmetrized
components, which had simpler structure and are more tractable
numerically. One practical way is to decompose systems w.f. into
so called FY components. Then FY equations often have obvious
advantage over Schr\"{o}dinger one, since it deals directly only
with FY components by avoiding construction of full system wave
function.

In that follows, we describe only four-particle FY equations; three-particle Faddeev equations
are self-contained in four-particle ones. 

\begin{figure}[h!]
\begin{center}\mbox{\epsfxsize=12.cm\epsffile{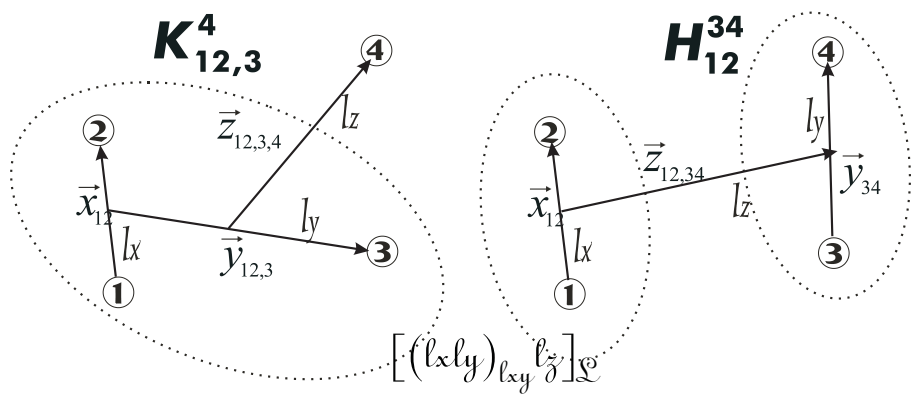}}\end{center}
\caption{Faddeev-Yakubovsky components $K$ and $H$.
Asymptotically, as $z\rightarrow \infty$, components $K$ describe
3+1 particle channels, whereas components $H$ contain asymptotic
states of 2+2 channels.}\label{Fig_4b_config}
\end{figure}

The calculations presented in this work were performed using the
framework of differential FY equations developed by S.P.\
Merkuriev and S.L.\ Yakovlev~\cite{Yakovlev,Merkuriev}. The major
step in these equations is the representation of the systems w.f. as a
sum of 18 components: 12 of type K, describing the asymptotic behavior
of different 3+1 particle channels, and 6 of type H, describing
the systems decomposition into two clusters of two particles:
\begin{equation}
\Psi =\sum\limits_{i<j}K_{ij,k}^{l}+\sum\limits_{i<j,k<l}H_{ij,kl}.
\label{Wave_func}
\end{equation}%
Here $\left( ijkl\right)$ indicates cyclic permutation of particle indices $(1234)$.
These FY components are coupled by 18 FY equations. Since all
particles are identical, several straightforward symmetry relations can be
established between different FY components. Using these relations, the
number of independent FY equations and components reduces to two:
\begin{eqnarray}
\left( E-\hat{H}_{0}-\hat{V}\right) K_{12,3}^{4} &=&\hat{V}(P^{+}+P^{-})%
\left[ (1+\varepsilon P_{34})K_{12,3}^{4}+H_{12,34}\right] \smallskip
\label{Fadd_Yak} \\
\left( E-\hat{H}_{0}-\hat{V}\right) H_{12,34} &=&\hat{V}\tilde{P}\left[
(1+\varepsilon P_{34})K_{12,3}^{4}+H_{12,34}\right] .  \notag
\end{eqnarray}%
Here $P^{+}=P_{12}P_{23}$, $P^{-}=P_{23}P_{12}$ and $\tilde{P}=P_{13}P_{24}$
are particle permutation operators.
 $\hat{H}_{0}$ is the kinetic energy operator of the system and $\hat{V}$
is the potential energy operator for the particle pair (12). The coefficient $%
\varepsilon $ is a Pauli factor: $\varepsilon =1$ for bosonic systems
and $\varepsilon =-1$ for systems of identical fermions.

By applying combination of permutation operators to the FY
components of Eq.~(\ref{Wave_func}), one can express the w.f. of the system
by means of two non-reducible FY components:
\begin{equation}
\Psi =\left[ 1+(1+P^{+}+P^{-})\varepsilon P_{34}\right]
(1+P^{+}+P^{-})K_{12,3}^{4}+(1+P^{+}+P^{-})(1+\tilde{P})H_{12,34}
\end{equation}

It is convenient to treat Eq.~(\ref{Fadd_Yak}) using relative
reduced coordinates. These coordinates are proportional to well known Jacobi
coordinates, which are simply scaled by the appropriate mass factors.
We use two different sets of relative reduced coordinates, defined as follows:
\begin{equation}
\begin{tabular}{ll}
$\overrightarrow{x_{ij}}$ & $=\sqrt{2\frac{m_{i}m_{j}}{m_{i}+m_{j}}}(%
\overrightarrow{r}_{j}-\overrightarrow{r}_{i})\smallskip $ \\
$\overrightarrow{y_{ij,k}}$ & $=\sqrt{2\frac{\left( m_{i}+m_{j}\right) m_{k}%
}{m_{i}+m_{j}+m_{k}}}(\overrightarrow{r}_{k}-\frac{m_{i}\overrightarrow{r}%
_{i}+m_{j}\overrightarrow{r}_{j}}{m_{i}+m_{j}})\smallskip $ \\
$\overrightarrow{z_{ijk,l}}$ & $=\sqrt{2\frac{\left(
m_{i}+m_{j}+m_{k}\right) m_{l}}{m_{i}+m_{j}+m_{k}+m_{l}}}(\overrightarrow{r}%
_{l}-\frac{m_{i}\overrightarrow{r}_{i}+m_{j}\overrightarrow{r}_{j}+m_{k}%
\overrightarrow{r}_{k}}{m_{i}+m_{j}+m_{k}})\smallskip $%
\end{tabular}%
\label{eq:J_coord}
\end{equation}%
for $K-$type FY components.  $m_{i}$ and $\overrightarrow{r}%
_{i}$ are respectively the $i$-th particle mass and position vector.
To describe $H-$type components, another set of coordinates is more appropriate:
\begin{equation}
\begin{tabular}{ll}
$\overrightarrow{x_{ij}}$ & $=\sqrt{2\frac{m_{i}m_{j}}{m_{i}+m_{j}}}(%
\overrightarrow{r}_{j}-\overrightarrow{r}_{i})\smallskip $ \\
$\overrightarrow{y_{kl}}$ & $=\sqrt{2\frac{m_{k}m_{l}}{m_{k}+m_{l}}}(%
\overrightarrow{r}_{l}-\overrightarrow{r}_{ki})\smallskip $ \\
$\overrightarrow{z_{ij,kl}}$ & $=\sqrt{2\frac{\left( m_{i}+m_{j}\right)
\left( m_{k}+m_{l}\right) }{m_{i}+m_{j}+m_{k}+m_{l}}}(\frac{m_{k}%
\overrightarrow{r}_{k}+m_{l}\overrightarrow{r}_{l}}{m_{k}+m_{l}}-\frac{m_{i}%
\overrightarrow{r}_{i}+m_{j}\overrightarrow{r}_{j}}{m_{i}+m_{j}})$%
\end{tabular}
\end{equation}

Using these coordinates, the kinetic energy operator is expressed
as a simple sum of Laplace operators:
\begin{equation}
H_{0}=-\frac{\hbar ^{2}}{2m}\Delta _{R}-\frac{\hbar ^{2}}{m}\left( \Delta
_{x}+\Delta _{y}+\Delta _{z}\right) .
\end{equation}%
Another big advantage of these coordinate sets is the fact that the degrees
of freedom due to the center of mass motion are separated.

The dimension of the problem can be further reduced by using the fact
that an isolated system conserves its total angular momentum
$\mathcal{J}$ and one of its projections $\mathcal{J}_{z}$. We
deal with systems of bosonic He atoms in their ground state, which
have a total spin equal to zero. In this case, the system w.f. is
independent of the spin and the orbital angular momentum is
conserved separately. Using this fact, we expand FY components
on a partial-wave basis (PWB) of orbital angular momentum
and, omitting the spin:

\begin{eqnarray}
\begin{tabular}{l}
$K_{ij,k}^{l}(\vec{x},\vec{y},\vec{z})=\sum\limits_{_{\alpha
}=(l_{x},l_{y},l_{xy},l_{z})}\frac{\mathcal{F}_{\alpha }^{K}(x,y,z)}{xyz}%
\left[ \left[ Y_{l_{x}}(\hat{x})\otimes Y_{l_{y}}(\hat{y})\right]
_{l_{xy}}\otimes Y_{l_{z}}(\hat{z})\right] _{\mathcal{LL}_{z}}$
 \\
$H_{ij,kl}(\vec{x},\vec{y},\vec{z})=\sum\limits_{_{\alpha
}=(l_{x},l_{y},l_{xy},l_{z})}\frac{\mathcal{F}_{\alpha }^{H}(x,y,z)}{xyz}%
\left[ \left[ Y_{l_{x}}(\hat{x})\otimes Y_{l_{y}}(\hat{y})\right]
_{l_{xy}}\otimes Y_{l_{z}}(\hat{z})\right] _{\mathcal{LL}_{z}}$
\end{tabular}
\label{FY_pwkh}
\end{eqnarray}

In this basis, the kinetic energy operator reads:
\begin{equation}
\hat{H}_{0}=\frac{\hbar ^{2}}{m}\left[ -\partial _{x}^{2}-\partial
_{y}^{2}-\partial _{z}^{2}+\frac{l_{x}(l_{x}+1)}{x^{2}}+\frac{l_{y}(l_{y}+1)%
}{y^{2}}+\frac{l_{z}(l_{z}+1)}{z^{2}}\right] .  \notag
\end{equation}

In Eq.~(\ref{FY_pwkh}), we have introduced the so called FY
amplitudes $\mathcal{F}_{\alpha }^{K}(x,y,z)$ and
$\mathcal{F}_{\alpha }^{H}(x,y,z)$, which are continuous functions
in radial variables $x$,$y$ and $z$. The symmetry properties of
the w.f. with respect to the exchange of two He atoms impose
additional constraints. One should have amplitudes only with an
even value of the angular momentum $l_{x}$, whether these
amplitudes are derived from $K$ or $H$ FY components.

In addition, for $H-$type amplitudes $\mathcal{F}_{\alpha
}^{H}(x,y,z)$, the angular momentum $l_{y}$ should be even as
well. The total parity of the system $\Pi$ is
 given by $\left( -\right) ^{l_{x}+l_{y}+l_{z}}$
independently of the coupling scheme ($K$ or $H$) used.

By projecting each of the Eq.~(\ref{Fadd_Yak}) to the PWB of
Eq.~(\ref{FY_pwkh}), a system of coupled integro-differential
equations is obtained. In general,  PWB is infinite and one
obtains thus an infinite number of coupled equations. This obliges
us, when solving these equations numerically, to make additional
truncations by considering only the most relevant amplitudes,
namely those which have the
smoothest angular dependency (small partial angular momentum values $%
l_{x},l_{y}$ and $l_{z}$).

\subsection{Boundary conditions}

Equations~(\ref{Fadd_Yak}) are not complete: they should be
supplemented with the appropriate boundary conditions for FY
components. It is usual to write boundary conditions in the
Dirichlet form, which at the origin should mean vanishing FY
components. However, the existence of a large strong repulsion
region (hard-core) corresponding to the inner part of He-He
potential brings additional complications. The relevant matrix
elements from the physical interaction region (shallow attractive
well) fade away in front of the huge repulsive hard-core terms,
thus resulting in severe numerical instabilities.

However, such a strong repulsion at the origin simply
indicates that two He atoms cannot get arbitrarily close to each other: for a repulsive region
of characteristic size $r_h$, the probability that two particles get closer to each other
at a certain distance $r=c<r_h$
will be vanishingly small. It means that the w.f. of the system
vanishes in part of a four-particle space, inside six multidimensional
surfaces $r_{ij}=c,$ where $r_{ij}$ is the distance between particle $i$ and $j$.
The most straightforward way to improve numerical stability would be to
avoid calculating the solution in at least part of the strong repulsion region,
and to impose by hand to the wave
function to be equal to zero in this region. Nevertheless, due to the complex geometry
of this domain in the nine-dimensional space of particle relative coordinates, this method
is not easy to put in practice.

A nice way to overcome this difficulty was proposed by Motovilov and Merkuriev~\cite{Motov}.
The authors showed that an infinitely repulsive interaction at $%
r_{ij}\leqslant c$, generates boundary conditions for the FY components
which can be ensured by setting:

\begin{gather}
\begin{tabular}{l}
$\left( E-\hat{H}_{0}-\hat{V}\right) K_{12,3}^{4}=0\smallskip $ \\
$\left( E-\hat{H}_{0}-\hat{V}\right) H_{12,34}=0$%
\end{tabular}%
\text{    for }x<c  \label{MM_BC}
\end{gather}
and
\begin{gather}
\begin{tabular}{l}
$K_{12,3}^{4}+(P^{+}+P^{-})\left[ (1+\varepsilon
P_{34})K_{12,3}^{4}+H_{12,34}\right] =0$ \\
$H_{12,34}+\tilde{P}\left[ (1+\varepsilon P_{34})K_{12,3}^{4}+H_{12,34}%
\right] =0$%
\end{tabular}%
\text{   for }x=c  \label{MM_BC1}
\end{gather}

In addition,  FY components asymptotic behavior should be
conditioned as well. For the bound state problem, the w.f. of the
system is compact,
therefore the regularity conditions can be completed by forcing the amplitudes $\mathcal{F}%
_{\alpha }^{K(H)}$ to vanish at the borders of the hypercube $\left[ 0,X_{\max }%
\right] \times \left[ 0,Y_{\max }\right] \times \left[ 0,Z_{\max }\right] $,
i.e.:
\begin{equation}
\mathcal{F}_{\alpha }^{K(H)}(X_{\max },y,z)=\mathcal{F}_{\alpha
}^{K(H)}(x,Y_{\max },z)=\mathcal{F}_{\alpha }^{K(H)}(x,y,Z_{\max })=0
\label{BC_BS}
\end{equation}

In the case of elastic atom-trimer (1+3) scattering, the asymptotic
behavior of the w.f. can be matched by simply imposing at
the numerical border $z=Z_{\max }$, the solution of the 3N bound state
problem for all the quantum numbers, corresponding to the open
channel $\alpha _{a}$. It worths reminding that only $K$-type
components contribute in describing 3+1 particle channels:
\begin{equation}
\mathcal{F}_{\alpha _{a}}^{K}(x,y,Z_{\max })=f_{\alpha _{a}}(x,y)
\label{BC_SS}
\end{equation}

Indeed, below the first inelastic threshold, at large values of $z$, the
solution of Eq.~(\ref{Fadd_Yak}) factorizes into a He trimer ground state
w.f. -- being solution of 3N Faddeev equations -- and a plane wave
propagating in $z$ direction with the momentum $k_{\alpha _{a}}=\sqrt{\frac{m%
}{\hbar ^{2}}\left( E_{cm}-E_{He_{3}}\right) }$. One has:
\begin{equation*}
\mathcal{F}_{\alpha _{a}}^{K}(x,y,z)\sim f_{\alpha _{a}}^{K}(x,y)\left[ \hat{%
\jmath}_{l_{z}}(k_{\alpha _{a}}z)+\tan (\delta )\hat{n}_{l_{z}}(k_{\alpha
_{a}}z)\right]
\end{equation*}

Here the functions $f_{\alpha _{a}}^{K}(x,y)$ are the Faddeev amplitudes
obtained after solving the corresponding He trimer bound state problem, whereas
$E_{He_{3}}$ is its ground state energy; $\hat{n}_{l_{z}}(k_{\alpha_{a}}z)$ and
$\hat{j}_{l_{z}}(k_{\alpha_{a}}z)$ are regularized Riccati-Bessel functions.
Equations (\ref{Fadd_Yak}) in
conjunction with the appropriate boundary conditions define the set of equations
to be solved. The numerical methods employed will be
briefly explained in the next subsection.

Once the integro-differential equations of the scattering problem are
solved, one has two different ways to obtain the scattering
observables. The easier one is to extract the scattering phases
directly from the tail of the solution, by calculating logarithmic
derivative ($\frac{\partial _{z}\mathcal{F}_{\alpha _{a}}^{K}(x,y,z)}{%
\mathcal{F}_{\alpha _{a}}^{K}(x,y,z)}$ ) of the open channel's $K$ amplitude $\alpha _{a}$ in
the asymptotic region:
\begin{equation}
\tan \delta = \frac{k_{\alpha _{a}}\hat{\jmath}_{l}^{\prime
}(k_{\alpha _{a}}z)-\frac{\partial _{z}\mathcal{F}_{\alpha _{a}}^{K}(x,y,z)}{%
\mathcal{F}_{\alpha _{a}}^{K}(x,y,z)}\hat{\jmath}_{l}(k_{\alpha _{a}}z)}{%
\frac{\partial _{z}\mathcal{F}_{\alpha _{a}}^{K}(x,y,z)}{\mathcal{F}_{\alpha
_{a}}^{K}(x,y,z)}\hat{n}_{l}(k_{\alpha _{a}}z)-k_{\alpha _{a}}\hat{n}%
_{l}^{\prime }(k_{\alpha _{a}}z)}
\label{log_der}
\end{equation}

This result can be independently verified by using an integral representation
of the phase shifts
\begin{equation}
k_{\alpha _{a}}\tan \delta =-\frac{m}{\hslash ^{2}}\int \Phi _{\alpha _{a}}^{(123)}\hat{%
\jmath}_{l}(k_{\alpha _{a}}z)(V_{14}+V_{24}+V_{34})\Psi dV.
\end{equation}%
$\Phi _{\alpha _{a}}^{(123)}$ is the trimer -- composed from the
He atoms indexed by 1,2 and 3 -- ground state w.f. normalized to
unity and $\Psi$ is normalized according to:
\begin{equation}
\Psi (\vec{x}_{i},\vec{y}_{i},\vec{z}_{i})=\Phi _{\alpha _{a}}^{(123)}\left(
\vec{x},\vec{y}\right) \left[ \hat{\jmath}_{l_{z}}(k_{\alpha _{a}}z)+\tan
(\delta )\hat{n}_{l_{z}}(k_{\alpha _{a}}z)\right] .
\end{equation}
Detailed discussions on this subject can be found in \cite{Fred_97,Thesis}.

\subsection{Numerical  methods }

In order to solve the set of integro-differential equations -- obtained when projecting
Eq.~(\ref{Fadd_Yak}) and the appropriate boundary conditions Eq.~(\ref{MM_BC}-\ref{BC_SS}) into
a partial wave basis --   the components $\mathcal{F}_{i}^{\alpha }$ are expanded in terms of piecewise
Hermite spline basis:
\[\mathcal{F}_{i}^{\alpha}(x,y,z)= \sum c^{\alpha}_{ijkl} S_j(x)S_k(y)S_l(z).\]
We use piecewise Hermite polynomials as a spline basis. In this way, the
integro-differential equations are converted into an equivalent
linear algebra problem with unknown spline expansion coefficients
to be determined. For bound states, the eigenvalue-eigenvector problem reads:
\begin{equation}
Ax=EBx,
\end{equation}%
where $A$ and $B$ are square matrices, while $E$ and $x$ are respectively
the unknown eigenvalue(s) and its eigenvector(s). In the case of the
elastic scattering problem, a system of linear algebra equations is obtained:
\begin{equation}
\left[ A-E_{cm}B\right] x=b
\end{equation}%
where $x$ is a vector of unknown spline expansion coefficients
and $b$ is an inhomogeneous term, generated when implementing the boundary conditions Eq.~(\ref%
{BC_SS}). For detailed discussions on the equations and  method
used to solve large scale linear algebra problems, one can
refer to~\cite{Thesis}.

\section{Trimer scattering and bound states\label{Sec:Trimer}}

As mentioned in the introduction, trimer states have been broadly
explored in many theoretical works. Hyperspherical, variational
and Faddeev techniques were used to calculate accurately bound
state
energies~\cite{Barletta,Sandhas,Roudnev_bs,Claude_mers,Sandhas_last,Fedorov_mers,Blume_mers}
as well as to test different He-He interaction models.
Nevertheless, we found  useful to consider these states as a first
step, before the more ambitious analysis of He tetramer states
could be undertaken. Special emphasis will be attributed to
He-He$_2$ scattering calculations, which are less studied and for
which some discrepancies were pointed
out~\cite{Roudnev_ss,Sandhas_last}.  Some arguments will also be
developed in favor of considering the first trimer excitation as
an Efimov state.

\bigskip
We present in Table~\ref{tab:trimer_conv} the convergence of the He trimer
states as a function of the partial-wave basis size.
It contains results for the ground ($B_3$) and first excited ($B_3^*$) state binding energies
as well as for the He-He$_2$ scattering length ($a_0^{(1+2)}$).
The basis truncations were made by limiting partial angular momentum
values, applying selection criteria $l_{x}=l_{y}\leq l_{max}$.
One can remark that the convergence is monotonic and quite similar
for the He$_{3}$  bound states and for the
He-He$_{2}$ scattering length calculations. These
results demonstrate that in order to ensure five digit accuracy, the
partial wave basis must include FY amplitudes
with angular momentum values up to 12. However, for reaching a
1\% accuracy,  $l_{max}=4$  turns out to be enough.
\begin{table}[h!]
\caption{\label{tab:trimer_conv}Convergence of He trimer calculations obtained when increasing
partial wave basis. In the three columns are respectively presented trimer
ground (B$_3$) and excited(B$^{*}_{3}$) state energies in mK, as well as atom-dimer scattering
length ($a_{0}^{(2+1)}$) in $\AA$.}
\begin{ruledtabular}
\begin{tabular}{c|ccc} 
$l_{max}$  & B$_3$ (mK) & B$^{*}_{3}$ (mK) & $a_{0}^{(1+2)}$ (\AA )   \\\hline 
      0     & 89.01      & 2.0093      &  155.39       \\
      2     & 120.67     & 2.2298      &  120.95       \\
      4     & 125.48     & 2.2622      &  116.37       \\
      6     & 126.20     & 2.2669      &  115.72       \\
      8     & 126.34     & 2.2677      &  115.61       \\
      10    & 126.37     & 2.2679      &  115.58 \\
      12    & 126.39     & 2.2680      &  115.56 \\
      14    & 126.39     & 2.2680      &  115.56 \\
\end{tabular}
\end{ruledtabular}
\end{table}

\bigskip
In Table~\ref{tab:3He_en} are summarized some relevant properties
of the trimer ground and excited states. Together with their
binding energies ($B$) we give some average quantities like
kinetic $<T>$ and potential energy $<V>$ values and moments of
interparticle distances $x_{ij}$. Our results are in perfect
agreement with other existing
calculations~\cite{Blume_mers,Barletta,Roudnev_bs}; the most
accurate of them have been included in this Table for comparison.
\begin{table}[h!]
\caption{\label{tab:3He_en}Mean values for He trimer ground and excited states.  In this
table B, T and V indicates respectively binding, mean kinetic and
potential energies calculated in mK;  $x_{ij}$ stands for interparticle distance.}
\begin{ruledtabular}
\begin{tabular}{r|cc|cc}  
 &\multicolumn{2}{c|}{Ground state}  & \multicolumn{2}{c}{Excited state} \\\hline 
B (mK)                                &  126.39 &126.4~\cite{Barletta}  & 2.268  & 2.265~\cite{Barletta}     \\
$<T>$ (mK)                               &   1658   &  1660~\cite{Barletta}     & 122.1   & 121.9~\cite{Barletta}      \\
$<V>$ (mK)                               &  -1785   &  -1787~\cite{Barletta}                         &   -124.5     & -124.2~\cite{Barletta}      \\
$\sqrt{<x_{ij}^2>}$ ($\AA$)           &   10.95  & 10.96~\cite{Barletta}    & 104.3        &  102.7~\cite{Roudnev_bs}\\
$<x_{ij}>$ ($\AA$)                   &   9.612  & 9.636~\cite{Barletta}    & 83.53        &  83.08~\cite{Barletta}           \\
$<x_{ij}^{-1}>$ ($\AA^{-1}$)            & 0.135    &                          & 0.0267       &         \\
$<x_{ij}^{-2}>$  ($\AA^{-2}$)        & 0.0230   & 0.0233~\cite{Kornilov} & 0.00218 &  \\

\end{tabular}
\end{ruledtabular}

\end{table}

\bigskip
The situation is more ambiguous for the He-He$_{2}$ scattering
length $a_0$. Results obtained by Sandhas et al.~\cite{Sandhas} --
solving Faddeev equations -- and Blume et al.~\cite{Blume_mers} --
using hypherspherical harmonics -- agreed with each other on a
scattering length value $a_{0}=126$ $\AA$. However
Roudnev~\cite{Roudnev_ss}, using also Faddeev equations, found a
smaller value $a_{0}=115.4\pm0.1$ $\AA$. It seems that in Sandhas
et al.~\cite{Sandhas} calculation a relatively small grid was
employed, which did not allow to disentangle the contribution of
virtual dimer break-up component in the asymptotic behavior of the
wave function. A more recent study~\cite{Sandhas_last} of the same
authors provided a value of $a_{0}=115.5\pm0.5$ $\AA$, in full
agreement with Roudnev's result.

\begin{figure}[h!]
\begin{center}
\mbox{\epsfxsize=12.cm\epsffile{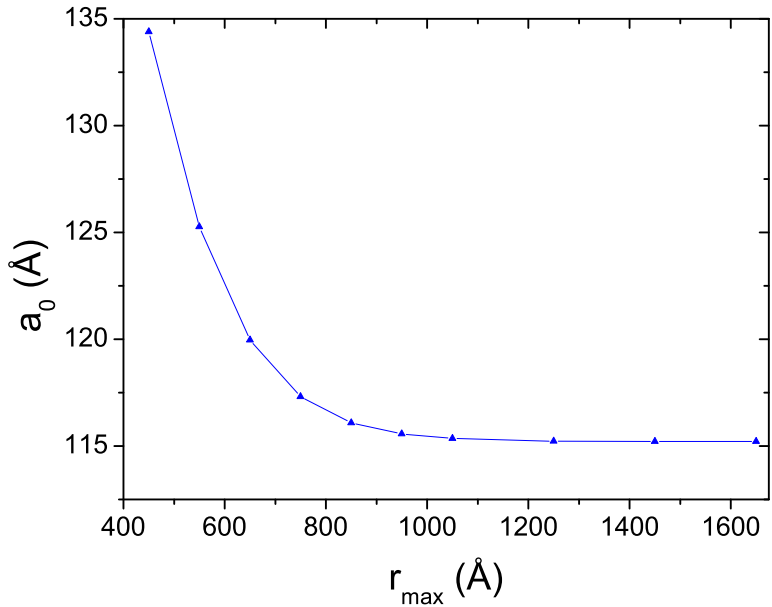}}
\end{center}
\vspace{-1.cm} \caption{"(Color online)" Convergence of
$He-(He)_{2}$ scattering length obtained when enlarging solution
domain of Faddeev equations in $He-(He)_{2}$ separation direction
(variable y) with additional discretization points.}
\label{Grid_conver}
\end{figure}

Our calculations are also very close to this value. When using
a numerical grid limited to $y_{max}=450\; \AA$ we obtain
 $a_{0}=134\; \AA$, still far from the final result.
Note that hyperradial grids employed in~\cite{Sandhas}  were limited
to $\rho _{\max }=\sqrt{x^{2}+y^{2}}$=460 $\AA$.
The results of
Table~\ref{tab:trimer_conv}  have been obtained using a grid with $y_{max}=950\; \AA$,
i.e. He-He$_2$ distance $r_{max}={\sqrt3\over2}y_{max}=823\; \AA$,
which is  large enough to reduce the grid-dependent variations to the fourth significant digit.
As a function of $r_{max}$, the
scattering length varies smoothly (see Figure~\ref{Grid_conver})
and converges towards the value $a_{0}(\infty )=115.2$ $\AA$,
very close to the one given by Roudnev $a_{0}=115.4\pm0.1$ $\AA$.

\bigskip
Whether or not the trimer first excitation is an
Efimov~\cite{Efimov} state has been an issue of strong
polemics~\cite{Gonsales,Esry_comm}. In fact, when using effective
range theory and describing system by zero-range interactions, it
is not difficult to show the appearance of an Efimov state.
However, the problem becomes complicated in calculations with
finite range potentials. Efimov states accumulate according to a
logarithmic law
\begin{equation*}
N\approx \frac{1}{\pi }\ln \frac{\left\vert a\right\vert }{r_{0}}
\end{equation*}%
and thus the two body scattering length should be each time
increased by a factor $\sim e^{\pi }\approx 23$ (or the dimer
binding energy reduced $\sim e^{2\pi }\approx530$ times) to allow
the appearance of an additional Efimov state. To handle this
property, the grids employed in calculations should be extremely
large and dense, capable on one hand  to accommodate extended wave
functions, and on the other hand to trace the very weak binding of
the third particle to the dimer. Such requirements can not be
fulfilled in the calculations with realistic interactions. Thus
the basic clue in claiming excited trimer to be an Efimov state is
the fact that this state disappears when the interatomic potential
is made less attractive~\cite{Gloeckle,Esry_init,Fedorov_non}.
Actually, if this potential is multiplied by an enhancement factor
$\gamma>$ 1 then the following effect is observed: first, the
difference between the trimer excited state binding energy
($B_3^*$) and the dimer one ($B_2$) increases with $\gamma$. Then,
for larger values of $\gamma$ the difference ($B_3^*-B_2$)
monotonously decreases and for $\gamma >$1.2 trimer excited state
moves below the dimer threshold and becomes a virtual one.

\begin{figure}[h!]
\begin{center}
\epsfxsize=12cm\mbox{\epsffile{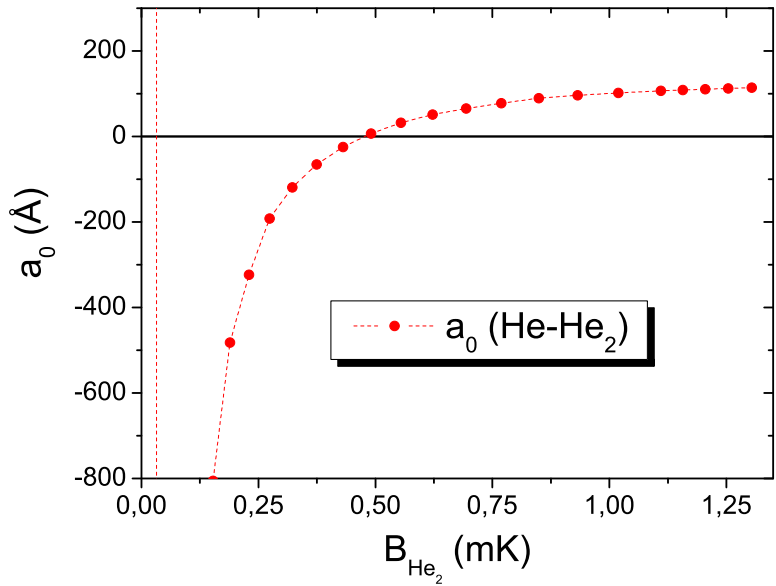}}\hspace{-0.6cm} %
\end{center}
\vspace{-1.2 cm} \caption{"(Color online)" The change of the
atom-dimer scattering length as a function of the dimer binding
energy.} \label{Fig_Efimov_st}
\end{figure}

The demonstration of Efimov effect can be accomplished only by
showing accumulation of new states when dimer binding energy
decreases. Such a demonstration has been given in
ref.~\cite{Kolg_Efim} using semi-empirical potential
HFD-B~\cite{Aziz2}. We would like to remark that the formation of new
states can be alternatively demonstrated by studying the $\gamma$-dependence
of the three-body scattering length, without the necessity of solving the bound state
problem. In scattering calculations, the numerical solution of 3-body
equations can be reduced to the interaction domain and the
wave function extended outside this region using analytical expressions~\cite{Jaume_dom_meth}.

In figure~\ref{Fig_Efimov_st} we display the behavior of the
He-He$_{2}$ scattering lengths, when the He-He potential is
multiplied by a scaling factor $\gamma <1$, according to
$\tilde{V}=\gamma V$. In this figure, the He-He$_{2}$ scattering
length is plotted as a function of the fictive dimer binding
energy. One can see that, when decreasing $\gamma $ -- scattering
length decreases. However in absence of  Efimov states one should
expect them increasing, since reducing $\gamma $,  the dimer
target becomes larger. Once $\gamma $ is reduced to $\approx
0.990$, the scattering length becomes negative and for values
$\gamma \approx 0.979 $ it exhibits a singularity going from
$a_{0}=-\infty $ to +$\infty$. This singularity corresponds to the
appearance of a new trimer bound state (i.e. second excited state
in He trimer). This analysis clearly demonstrates that He trimer
excited state is an Efimov one. It is worth mentioning that for an
enhancement factor $\gamma =0.979$, the He dimer binding energy is
only 0.046 $mK$.

\section{\label{Sec:tetramer}Tetramer states}

The major aim of this article is to provide a comprehensive
analysis of the four-atomic He compound (tetramer). The first
efforts to describe this system were made already in the late
seventies by S. Nakaichi et al.~\cite{Maeda_var} using a
variational method. Latter on, variational Montecarlo techniques
were used by several
authors~\cite{Blume_mers,Bresanini,Lewerenz,Guardiolla} to compute
the tetramer ground state. These methods are very powerful in
calculating $L^{\Pi}=0^+$ bound state properties, but are seldom
generalized to describe excited states and are not appropriate for
scattering problem.

FY techniques were also used by S. Nakaichi et al.~\cite{Maeda_FY}  to
calculate the tetramer ground state binding energy and the He-He$_3$
scattering length.
However in order to reduce the -- at that time -- outmatching numerical costs,
some important approximations were made.
The He-He potential was restricted to S-wave and written as a one-rank
separable expansion and the same expansion was used to
represent the FY amplitudes. These approximations led to a tetramer
ground state which is underbound by 40\% with respect to their own variational result
~\cite{Maeda_var}.
A  recent attempt to calculate the He tetramer binding energy
using S-wave FY equations was done in~\cite{Roudnev_He4},
although without separable expansion of FY amplitudes.

\begin{table}[h!]
\caption{\label{tab:tetramer_conv}Convergence of the He tetramer
calculations obtained when increasing the partial wave basis of Eq.~(\ref{FY_pwkh}).
The two columns  represent respectively the tetramer ground state binding energy in mK (B$_4$)
the and the atom-trimer scattering length ($a_{0}^{(3+1)}$) in $\AA$.}
\begin{ruledtabular}
\begin{tabular}{lll}
$\max (lx,ly,lz)$ & B$_{4}$ (mK) & $a_{0}^{(3+1)}$ (\AA ) \\\hline 
0 & 348.8 &  $\approx$-855 \\
2 & 505.9 &  190.6\\    %
4 & 548.6 &  111.6 \\   %
6 & 556.0 &  105.9 \\   %
8 & 557.7 &  103.7 \\  %
\end{tabular}
\end{ruledtabular}
\end{table}

The FY calculations we present here contain no approximation other
than the finite basis set used in the partial wave expansion
(\ref{FY_pwkh}). This basis set included amplitudes with internal
angular momentum not exceeding a given fixed value $l_{max}$, i.e.
fulfilling the condition  $max(l_x,l_y,l_z)\leq l_{max}$. The
largest basis we have considered has $l_{max}=$8, and consist of
180 FY amplitudes, a number by two order of magnitude larger than
in preceding calculations. Note, that the smallest basis, which is
often referred to as S-wave approximation, is obtained by fixing
$l_{max}=0$ and requires only 2 amplitudes, one of type K and one
of type H. The convergence is displayed in
Table~\ref{tab:tetramer_conv} for the tetramer ground state
binding energy and He-He$_3$ scattering length.

\begin{figure}[h!]
\begin{center}
\epsfxsize=12cm\mbox{\epsffile{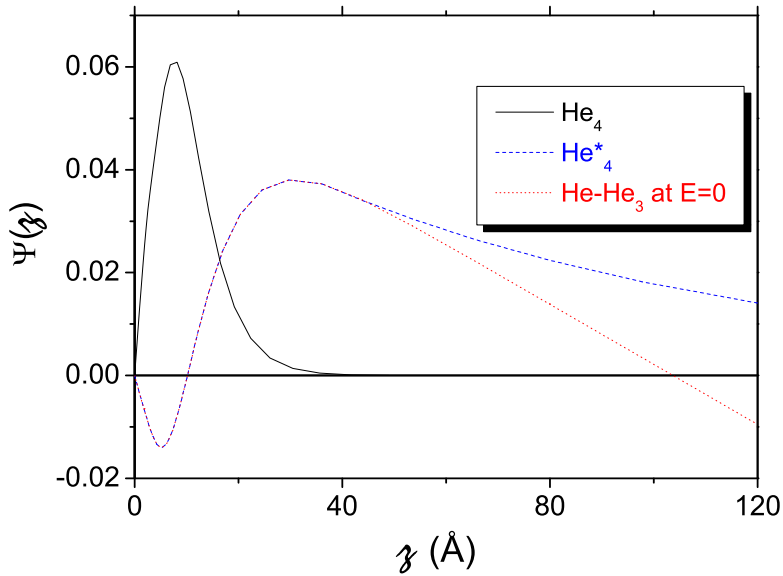}}\hspace{-0.6cm} %
\end{center}
\par
\vspace{-1.2 cm} \caption{"(Color online)" Comparison of the
functional dependence of K-type FY components in one He atom
separation from He$_3$ core direction. Single, dashed and dot line
curves correspond respectively to tetramer ground, excited state and
He-He$_3$ zero energy scattering wave functions.}
\label{Fig_tetramer_wf}
\end{figure}

The corresponding FY K-amplitudes are displayed in Figure~\ref{Fig_tetramer_wf}
as a function of the He-He$_3$ distance $r=\sqrt{2\over3} z$. One can see the different
scales involved in the bound state and zero energy scattering wave function.

The convergence of the tetramer calculations is sensibly slower
than the one observed for the trimer case (see Table~\ref{tab:trimer_conv}).
Such a deterioration is due to the complex structure of the involved FY components.
Indeed, each He atom pair brings an additional hard-core region;
the ensemble of these regions crosses over in the multidimensional configuration space
\footnote{In two particle case one
has single hard-core region in three-dimensional particle space;
this region has simply a form of the sphere of radius $r_{hc}$
placed at the origin. One has three hard-core conditions for three
particle system, these regions are intersecting in six-dimensional
multiparticle space. Four particle case already has six hard-core
conditions for nine-dimensional multiparticle space.}
and results into a single domain with non trivial geometry.
Inside this multidimensional domain, the total wave function
must vanish by cancelling the contributions of the different FY amplitudes,
what can be achieved only at the price of increasing its functional complexity.

The convergence of the He-He$_3$ scattering calculations is also slightly
slower than for tetramer ground state. This is a consequence of the
scattering wave function, which presents a richer structure than the ground state one,
as can be seen in Figure~\ref{Fig_tetramer_wf}.
Tetramer ground state has a rather simple geometry:
the four He atoms minimize their total energy by forming a tetrahedron.
The total energy of the 1+3 scattering state ($E\approx-126$ mK) is
considerably larger than the ground state one ($E\approx-558$
mK) and provides more flexibility to each atom.
Furthermore, since the scattering length
value is much larger than the size of the trimer  target,
the scattering wave function turns to be strongly asymmetric.

Despite these difficulties, the PW basis we
have used enables to reach rather accurate results: the tetramer ground
state binding energy is converged up to 0.4\%, while the final variation
of the scattering length does not exceed 2\%.
Our best result for the ground state binding energy of He$_4$, B=558 mK,
is in perfect agreement with the B=559(1) mK value, provided by several
variational Monte-Carlo techniques~\cite{Blume_mers,Bresanini,Lewerenz}.

\bigskip
It is interesting to compare the bound state result with the
effective field theory (EFT) predictions.
It follows from this approach that systems governed by large scattering lengths
should exhibit universal properties. The wave functions of such systems have
very large extensions with only a negligible part located inside
the interaction region.
The detailed form of the short-range potential does not matter,
since the system probes it only globally.
It should be therefore
possible to describe its action using only few independent parameters
(physical scales). One scale is obviously required,
 to fix the two-body binding energy or the scattering length.
Keeping fixed the two-boson binding energy, the
three bosons system would collapses if the interaction range
tends to zero, a collapse known as the Thomas effect~\cite{Thomas}.
This indicates that three-body system is
sensitive to an additional scale,  which can be
determined by fixing one three-body observable (for instance the 3-particle
binding energy or the 1+2 scattering length)~\cite{Tobias,Fedorov}.

It seems~\cite{Platter} that the four-boson binding energies
remains finite if none of its three-boson subsystems collapse.
Furthermore four- and three-body binding energies are found to be
correlated~\cite{Platter}. The simplest way to establish such a
correlation law is by using contact interactions. In this way, the
two-body binding energy is fixed by the parameter strength of
zero-range two-body force, whereas the three-body collapse is
avoided and its binding energy is fixed by introducing repulsive
three-body contact term. Using this model Platter et
al.~\cite{Platter} demonstrated that inside quite a large domain
of dimer-trimer binding energy ratio ($B_2/B_3$), the correlation
between tetramer and trimer binding energies is almost linear. In
the same study, a numerical approximation of this correlation law
was obtained.

If we take our dimers
($B_{He_2}=1.30348$ mK) and trimers binding energies as the scales for
EFT with contact interaction, one obtains a tetramer binding energy
of 485 and 491 mK respectively, depending on which trimer energy --
ground or excited -- is used. This result is only
by 13\% smaller than our most accurate result.

One should remark that the EFT scaling laws are derived using
contact interactions, which act only in S-wave. Therefore it would
be more consistent to compare our values obtained using S-wave
approximation ($l_{max}=0$). In this case, the EFT formulas give
respectively $B_{He_4}=330$ mK and 325 mK depending on which
trimer fixes the scale. These values must be compared with our
result $B_{He_4}=348.8$ mK, i.e.  EFT is only  by 6\% off.

The good agreement between exact and EFT results, proves the great prediction power of the
last approach. EFT works surprisingly well even beyond its natural
limit of applicability: systems with a size significantly
exceeding the range of interaction. Note, indeed, that the He tetramer and
trimer ground states are rather compact objects, in which
the interatomic separation is about $\approx10 \;\AA$, and therefore comparable
with He-He interaction range $\sim 3 \;\AA$.

\begin{table}[h!]
\caption{\label{tab:tetramer_prop}Mean values for He tetramer ground and excited states.  In this
table B, T and V indicates respectively binding, mean kinetic and
potential energies calculated in mK;  $x_{ij}$ stands for interparticle distance.}
\begin{ruledtabular}
\begin{tabular}{r|cc|cc}
 &\multicolumn{2}{c|}{Ground state}  & \multicolumn{2}{c}{Excited state} \\\hline
B (mK)                               &   557.7     & 559~\cite{Lewerenz}          & 127.5 & 128-130~\cite{Platter}     \\
T (mK)                               &   4107           &                         & 1900  &     \\
V (mK)                               &   -4665          & -4850~\cite{Lewerenz}     & -1913      &       \\
$\sqrt{<x_{ij}^2>}$ ($\AA$)          &  8.40           &                          & 34.4  &  \\
$<x_{ij}>$ ($\AA$)                   &  7.69           & 7.76~\cite{Lewerenz}     & 24.8 &            \\
$<x_{ij}^{-1}>$ ($\AA^{-1}$)         &  0.156     &                              & 0.088 &    \\
$<x_{ij}^{-2}>$  ($\AA^{-2}$)        &  0.0286     & 0.0251~\cite{Kornilov}     & 0.013 &
\end{tabular}
\end{ruledtabular}
\end{table}

\bigskip
Let us discuss now the He-He$_3$ scattering results.
Nakaichi et al.~\cite{Maeda_FY}, using S-wave
separable expansion of the outdated HFDHE2 potential,
obtained a large and negative scattering length a$_0=-$116 $\AA$.
Using the same HFDHE2 potential and restricting to $l_{max}$=0 amplitudes we have got
also a negative, although much larger, scattering length value a$_0\approx$-5600 $\AA$.
This result is however very unstable, as a consequence of the $l_{max}$=0 basis
inability to describe the non-trivial behavior of the FY components in the hard-core region.
If we add  $\max(l_y,l_z)\leq$4  amplitudes, the scattering length reduces to $\approx$-898 $\AA$
for the same HFDHE2 interaction model limited to S-wave.
Similar calculations with the LM2M2 potential give a$_0\approx$-450 $\AA$.
As one can see in Table~\ref{tab:tetramer_conv}, the full potential
must be used in order to obtain converged results.
The presence of He-He interaction in
higher partial waves cardinally changes the physics of He tetramer: not only the size but also
the sign of the resonant scattering length changes, thus indicating the emergence of
a new excited state in this compound.

Another effort to evaluate He-He$_3$ scattering length was made by
Blume et al.~\cite{Blume_mers}. In their work, effective $He-He_n$
potentials were constructed starting from the same LM2M2 He-He
interaction model. Without taking into account the particle
correlations, Blume et al. provided a positive He-He$_3$
scattering length a$_0$=56.1 $\AA$.

\bigskip
As already mentioned, the large positive scattering length
indicates the presence of a tetramer excitation close to the trimer
ground state threshold.
Much physics about this tetramer state can be learned by studying
the behavior of the zero energy 1+3 scattering w.f., or what is even more
practical, its FY components. These components
are only partially symmetrized and have a more transparent asymptotic
behavior~\cite{Merkuriev,Yak_Merk_assympt}.
The structure of K-type FY component displayed in Figure \ref{Fig_tetramer_wf},
proves that this state is the first tetramer excited state: the
corresponding open channel FY amplitudes have two nodes in z, the
He-He$_{3}$ separation direction. The first node is situated at
$\approx$10 $\AA$, i.e. inside the He$_{3}$ cluster, indicating
the presence of a compact ground state. The second
node is situated at $\approx$103.7 $\AA$, which
coincides with the scattering length value. This coincidence is not
accidental: at such He-He$_3$ separations, the single He atom is already out of
the interaction domain of the He$_{3}$ cluster. Close to this node, the FY
components reach the well known linear behavior:
\begin{equation}
\mathcal{F}_{\alpha _{a}}^{K}(x,y,z)\sim \phi_{He_3}(x,y)(\sqrt{\frac{2}{3}}z-a_0)
\label{eq:hehe3assymp}
\end{equation}
The factor $\sqrt{\frac{2}{3}}$ in front of the z-coordinate is
due to mass scaling of Jacobi variables Eq.~(\ref{eq:J_coord}).

A direct  calculation of the He$_{4}$ excited state  represents
nowadays a hardly realizable numerical task. This state is very
weakly bound and its wave function very extended. In order to
numerically reproduce it, one is obliged to use a very large and
dense grid. On the other hand, one should be able to ensure a high
accuracy to trace a small binding energy difference. For the time
being, it constitutes an insurmountable obstacle in the context of
the large computing power demanding four-body calculations.

Nevertheless the vicinity of tetramers excited state to the He-He$_{3}$ continuum makes
possible the extraction of its binding energy from the scattering results.
A bound state is identified with a S-matrix $S_{l}(k)$ pole on the positive imaginary
axis of the complex momentum plane.
Since $S_{l}(k)$ is unitary, its general form close to the pole $k=ik_{0}$ is:
\begin{equation}
S_{l}(k)=\frac{k+ik_{0}}{k-ik_{0}}=e^{2i\delta }.
\label{sl_matrix}
\end{equation}
The momentum $k_{0}$ is related to the tetramer binding energy measured with respect to trimer ground state threshold by $\Delta B_{0}=\frac{\hbar ^{2}k_{0}^{2}}{2\mu_{3+1}}$, where
$\mu_{3+1}=3m/4$ is the reduced atom-trimer mass.

On the other hand, the well known effective range expansion can be used to approximate the low energy phase shifts:
\begin{equation}
k^{2l+1}\cdot {\rm ctg}\delta
=-\frac{1}{a_{l}}+\frac{1}{2}r_{l}k^{2}+o(k^{2})
\label{Power_exp}
\end{equation}
Combining relations (\ref{sl_matrix}) and  (\ref{Power_exp}), one obtains an expression
for the bound state momentum $k_{0}$ in terms of the low energy parameters. For
the  He-He$_{3}$ scattering states with $l_z=0$,  it reads:
\begin{equation}
\frac{1}{2}r_{0}k_{0}^{2}-k_{0}+\frac{1}{a_{0}}=0\label{Bnd_eng}
\end{equation}
\vspace{-0.95 cm}
\begin{figure}[h!]
\begin{center}
\epsfxsize=12cm\mbox{\epsffile{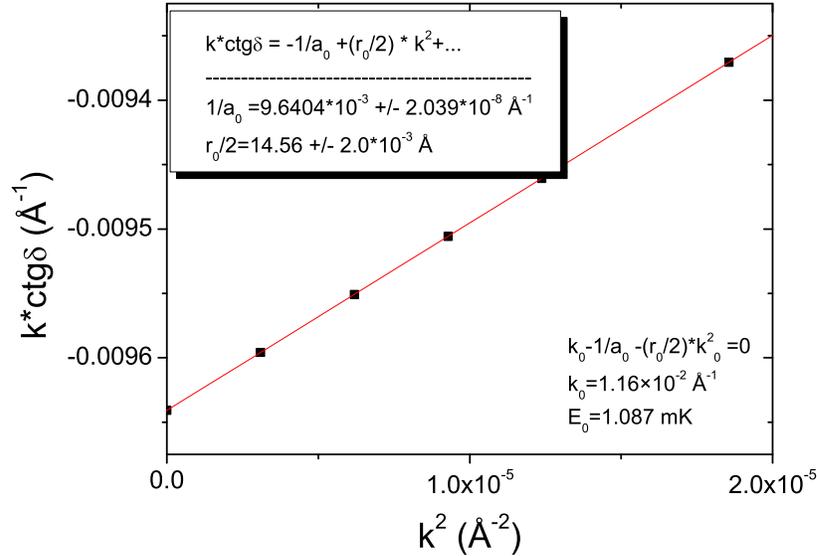}}\hspace{-0.6cm} %
\vspace{-1. cm} \caption{"(Color online)" Extrapolation of He
tetramer excited state binding energy from the He-He$_3$
scattering calculations. Low energy phase shift fit is made for
k*ctg$\delta$ values as a linear function of momentum squared
(k$^2$).} \label{Fig_extrap_tetra}
\end{center}
\end{figure}

It follows from Eq.~(\ref{Bnd_eng}) that in order to obtain
the unknown binding energy it is sufficient to know the low energy coefficients $a_{0}$ and $r_{0}$.
The scattering length  value $a_{0}$ has been already determined
and the effective range $r_{0}$ can be extracted
by fitting with Eq.~(\ref{Power_exp}) the He-He$_3$ low energies phase shifts.
The corresponding extrapolation procedure is illustrated in
Fig.~\ref{Fig_extrap_tetra}. One can see that inside the
considered momentum region,  close to $k=|k_0|$, the effective
range expansion works perfectly
well. The obtained binding energy, $\Delta B_{0}=1.087 mK$,
should not suffer much from  higher order momentum
terms in the expansion (\ref{Power_exp}).
Nevertheless, the He-He$_3$ effective range
is pretty large $r_{0}$=29.1 $\AA$ and influences
significantly the extrapolated binding energy value.
By ignoring this term, one would get
\begin{equation}
\Delta B^{(0)}_{0}=\frac{\hbar ^{2}}{2\mu_{3+1}a_{0}^{2}},
\end{equation}
and a binding energy of only $\Delta B^{(0)}_{0}=0.751 mK$ will be obtained, i.e.
a value by 30\% smaller.


We  finally predict the existence of a $L^{\Pi}=0^+$  tetramer excited state with binding energy
$B^{*}_{He_4}=\Delta B_0+B_{He_3}=127.5$ mK.
This value compares
well with the EFT prediction~\cite{Platter} discussed above, which
gives the range $B^{*}_{He_4}\in[128-130]$  mK, a dispersion due to the fact that the interpolation
was done for the total
binding energy and not the sensibly smaller relative value $\Delta B_0$.

\begin{figure}[h!]
\vspace{-1. cm}
\begin{center}
\epsfxsize=14cm\mbox{\epsffile{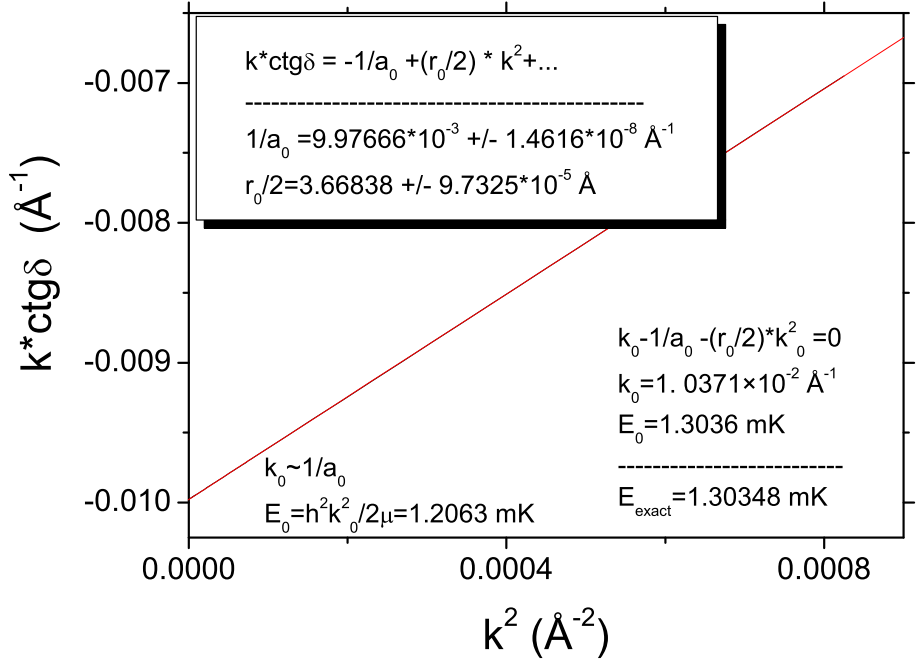}}\hspace{-0.6cm} %
\end{center}
\vspace{-1. cm} \caption{"(Color online)" He dimer binding energy
extrapolated from He-He scattering calculations. Low energy phase
shift k*ctg$\delta$ is fitted with a linear function of $k^2$.
Exact k*ctg$\delta$(k$^2$) values are indistinguishable from the
linear fit.}\label{Fig_extrap_dimer}
\end{figure}

The validity of the procedure we use to obtain the binding energy of the excited tetramer
can be verified in the dimer case,  for which direct bound
state calculations causes no difficulty.
The procedure is illustrated in Fig.~\ref{Fig_extrap_dimer}.
The accurate dimer binding energy is
$B_{He_2}=1.30348 $ mK, a value very close to $\Delta B_0=1.087$ mK, and allows
to control the inaccuracy made when disregarding higher order
 terms in expansion (\ref{Power_exp}). By considering only two terms in the
expansion~(\ref{Power_exp}) we have got $B_{He_2}=1.3036$ mK, only
differing in the fifth significant digit from the directly calculated
value. In addition, the He-He potential effective range extracted
by fitting the low energy phase shift is  $r_{0}$=7.337 $\AA$.
This value can be independently calculated applying the Wigner
formulae  from zero-energy scattering wave function, which gives
$r_{0}$=7.326 $\AA$, in nice agreement with the extrapolated one.
Such an agreement demonstrates the validity of our approach.

One should however mention that the very same procedure is not applicable to the trimer
excited state calculation.
If done, it would lead to a complex momentum $k_{0}=(1.3\pm0.8i)\cdot10^{-2}$ $\AA^{-1}$.
This is a consequence of a dimer binding energy  almost as small as $\Delta
B^{*}_{He_{3}}=B^{*}_{He_3}-B_{He_2}$. In this case, the scattering
phase shifts are affected by two open thresholds and thus a single
channel S-matrix theory is not appropriate. In He-He$_3$
scattering, the nearest threshold He$_2$-He$_2$ opens only for
scattering energies E$_{cm}\approx$124 mK and thus is well
separated from the energy region of interest ($\sim$ 1 mK).

\bigskip
It is interesting to compare the effective ranges for He-He,
He-He$_2$ and He-He$_3$ systems.  They are respectively
$r_0$=7.33, 79.0 and 29.1 $\AA$. It is not surprising that the
atom-dimer effective range is the largest one: this is a
consequence of a dimer ground state which is the most extended of
the three considered targets. The atom-trimer effective range is
more than a third of the atom-dimer one and is significantly
larger than the range of the He-He potential. This suggest that
the trimer ground state has a structure with a  sizeable
probability to find a single He atoms separated by 20-30 $\AA$
apart from its center.

\bigskip
We can still use the He-He$_{3}$ scattering w.f. to obtain a
relatively good description of the tetramer excited state. In
fact, the left hand side of FY equations \ref{Fadd_Yak} for the
bound and zero-energy He-He$_{3}$ state differ only by a small
energy term, which has a little effect on the w.f. inside the
interaction region, dominated by large kinetic and potential
terms. These functions sensibly differ only in the He-He$_3$
asymptotes, where they can be described
analytically~\cite{Merkuriev,Yak_Merk_assympt} as a tensor product
of a strongly bound trimer in its ground state and a plane wave of
the remaining He atom with energy E-E$_{He_3}$. In practice, the
two w.f. differ only by FY amplitudes contributing to the open
channels. For zero-energy scattering, one has a linearly diverging
open channel FY amplitude as described in
Eq.~(\ref{eq:hehe3assymp}). For bound state these amplitudes
converge very slowly with a small exponential factor:
\begin{equation}
\mathcal{F}_{\alpha _{a}}^{K}(x,y,z)\sim \phi_{He_3}(x,y) e^{-\sqrt{\frac{2}{3}}k_0*z}
\label{eq:he4assymp}
\end{equation}
The closed channel FY amplitudes rapidly vanish being shrunken by
a relatively large $e^{-\sqrt{\frac{m*B_{He_3}}{\hbar^2}}*\rho}$
exponent, with $\rho=\sqrt{x^2+y^2+z^2}$.

Using this approximation for the tetramer excited state  wave function,
we have calculated some of its properties, which are summarized in Table~\ref{tab:3He_en}.

\textit{\bigskip }

\section{\label{Sec:rotational}Tetramer and trimer rotational states}

\begin{figure}[h!]
\begin{center}
\epsfxsize=6cm\mbox{\epsffile{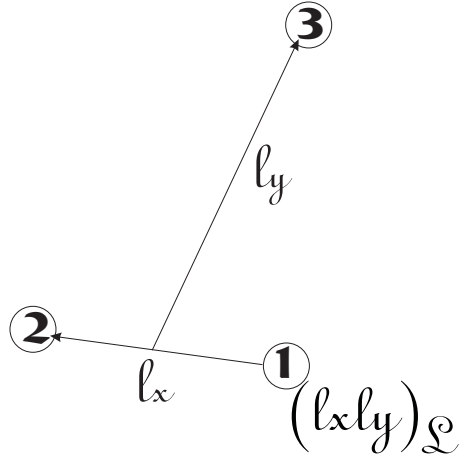}} %
\end{center}
\caption{Coupling scheme for three-particle FY component.}
\label{Fig_trimer_cpl}
\end{figure}

Considering that $^{4}$He is a spinless atom, the rotational
states of $^{4}$He-multimers can be classified according to their
parity ($\Pi$) and total orbital angular momentum ($L$). FY
components are very useful to classify these states as well as to
analyse their properties. As we have mentioned in
section~\ref{Sec:Theory}, the $l_{x}$ quantum number  must be even
(see Fig.~\ref{Fig_trimer_cpl}). It follows that for He trimer,
the $L^{\Pi}=0^{-}$ states are forbidden.  Further we classify the
He trimer states into those which can break into dimer and a
freely propagating He atom, and those in which dimer states are
forbidden by symmetry requirements. In the first group, we have
states with $L^{\Pi}=(1^{-},2^{+},3^{-},...)$. To be decay stable,
these states should be more bound than dimer. The other group is
formed by the $L^{\Pi}=(1^{+},2^{-},3^{+},...)$ states, which can
only break into three helium atoms.

Clearly, the $1^{-}$ is the most promising candidate for a trimer
stable rotational state. This state is realized with the smallest
partial angular momentum, i.e. $l_{x}+l_{y}\geq1$, and
consequently contains the smaller centrifugal terms in the
Hamiltonian. The non-existence of this state would imply the
non-existence of other states of the same decay group -- i.e.
$L^{\Pi}=(1^{-},2^{+},3^{-},...$) -- since they involve larger
centrifugal energies: $l_{x}+l_{y}\geq2$. The existence of stable
trimers in the second group is very doubtful; the most favorable
would be the $2^{-}$ one, for which one has already
$l_{x}+l_{y}\geq3$.

\bigskip
In a similar way, we can classify the rotational states of He
tetramer (see Fig.~\ref{Fig_4b_config}). In the first group, we
have states $L^{\Pi}= (1^{-},2^{+},3^{-},..)$ for which the first
decay threshold is the trimer ground state. The most promising
state inside this group is the $1^{-}$ one. For this state, the
condition $l_{x}+l_{y}+l_{z}\geq1$ must be satisfied.

In the second group, we have trimer decay stable states $L^{\Pi}=
(1^{+},2^{-},3^{+},..).$ These states can neither break into
trimer-atom pair nor into two dimers: their reference threshold is
dimer plus two free atoms. $L^{\Pi}=0^{-}$ state represents a
special case: it can be broken only into four free atoms.

The FY formalism we are using allows  to calculate multiparticle
rotational states with the same ease as zero angular momentum
ones. However, none of trimer or tetramer rotational states have
been found stable. The non-existence of weakly bound states can be
easily verified using low energy scattering techniques for trimer
and tetramer states in the first decay groups (two cluster). In
addition, the calculated scattering length turns out to be an
indicator for the strength of the interaction between the
scattered clusters and thus tests if nearthreshold bound or
resonant states are present.

\bigskip
For atom-dimer scattering in 1$^{-}$, one obtains a large positive
scattering length $(a_1)^{\frac{1}{3}}$=114.2 $\AA$. If He-He
interaction is changed, this length scales with the dimer size. If
we add a small attractive three-body interaction to force a trimer
binding without affecting dimer, this scattering length becomes
smaller. In fact, when the attractive three-body force is weak,
this scattering length reduces very slowly. Only when this
additional force becomes rather strong, close to the critical
value binding 1$^{-}$ trimer, the scattering length falls down to
-$\infty$. It crosses a singularity, passing from -$\infty$ to
+$\infty$ at the value where 1$^{-}$ trimer bound state appears,
and then again stabilizes to large positive value. This clearly
indicates that no trimer rotational state exist with quantum
numbers  $L^{\Pi}=(1^{-},2^{+},3^{-}...)$. An even much stronger
attractive three-body force is required to bind the second group
$L^{\Pi}=(1^{+},2^{-},3^{+},...)$ states, thus excluding the
possible existence of bound or even nearthreshold resonant He
trimer states.

The case of rotational tetramers is identical to the trimers one.
A positive scattering length $(a_1)^{\frac{1}{3}}\approx$12.28
$\AA$ is also  obtained for He-He$_3$ scattering with
$L^{\Pi}=1^{-}$. This value is sensibly smaller than the one for
$1^{-}$ atom-dimer scattering, which simply results from the
scaling with the target size. As in trimers case, one has to apply
a strong additional attractive force to reduce this scattering
length  to -$\infty$, i.e. to force the $1^{-}$ tetramers binding.
Tetramers in the second decay group, as well as $L^{\Pi}=0^{-}$,
seem to be even less favorable: they require an even stronger
additional force to be bound. We conclude therefore that no bound
or even nearthreshold resonant states should exist for tetramers
with $L^{\Pi}\neq0^{+}$.

\section{\label{Sec:Conclusions}Summary}

In this paper, we have outlined Faddeev-Yakubovski equation
formalism in configuration space. It enables a consistent
description of bound and scattering states in multiparticle
systems. This formalism was applied to study the lightest
($N=2,3,4$) systems of He atoms using realistic He-He interaction.
We have presented accurate calculations for bound He trimer and
tetramer states, as well as for low energy atom-dimer and
atom-trimer scattering.

\bigskip
Our main results concern the He tetramer states.
We have obtained a tetramer ground
state binding energy of $B=558$ mK, in perfect agreement with the most accurate variational results.

\bigskip
The first  realistic calculation of He-He$_3$ scattering length has been achieved,
with the prediction $a_0$=104 $\AA$. Such a large value indicates the existence of
a tetramer excited state close to the trimer ground state threshold.

Its binding energy can hardly be determined in a direct bound
state calculation, due to the difficulties in accommodating a very
extended wave function. We have shown that this energy can be
obtained by applying an effective range expansion to low energy 1+3
scattering states. We predict the existence of a $L^{\Pi}=0^{+}$
He tetramer excited state with a binding energy of 127.5 mK,
situated 1.09 mK below the trimer ground state.

\bigskip
Finally, we have studied the possible existence of rotational states in three and four atomic He compounds.
It has been shown that neither the He trimer nor the tetramer have bound rotational ($L^{\Pi}\neq0^{+}$)  states.
The existence of corresponding nearthreshold resonances is also doubtful.

\bigskip
{\bf Acknowledgements:}
Numerical calculations were performed at
Institut du D\'eveloppement et des Ressources en Informatique Scientifique (IDRIS) from  CNRS
and at Centre de Calcul Recherche et Technologie (CCRT) from CEA Bruy\`eres le Ch\^atel.
We are grateful to the staff members of these two organizations for their kind hospitality and useful advices.

\end{document}